\documentclass[aps,prb,reprint,groupedaddress,amsfonts,amssymb,amsmath]{revtex4-2}

\usepackage{lineno,hyperref,amsmath,graphicx}

\begin{document}
\title{Critical temperatures of a model cuprate}

\author{Yu. D. Panov}

\email[]{yuri.panov@urfu.ru}

\affiliation{Ural Federal University, Ekaterinburg, Russia}

\begin{abstract}
The problem of competing orderings in the high-temperature cuprate materials is widely discussed for the last years. 
We present the mean-field approximation results for the spin-pseudospin model accounting for the on-site and inter-site correlations, the antiferromagnetic exchange coupling, the one- and two-particle transport. 
The explicit form of the equations for the critical temperatures of the most significant order parameters of the model are given.
\end{abstract}
\maketitle

\section{Introduction}

Over the past fifteen years, numerous experimental results have shown the presence of various ordered states in the so-called pseudogap region of the cuprate phase diagram.
The relationship between superconductivity and other competing orders remains a hotly debated topic in the physics of high-$T_c$ cuprate materials~\cite{Fradkin2012}.
A minimal model to describe the charge degree of freedom in cuprates was introduced recently 
and makes use of the $S=1$ pseudospin formalism \cite{Moskvin2011,Moskvin2013}. 
It implies that for the CuO$_4$ centers in CuO$_2$ plane
the on-site Hilbert space reduced to a charge triplet 
formed by the three many-electron valence states CuO$_4^{7-,6-,5-}$ (nominally Cu$^{1+, 2+, 3+}$). 
These states are described as the components of the $S=1$ pseudospin triplet with $M_S = {-}1,\, 0,\, {+}1$.
Effective pseudospin Hamiltonian of the model cuprate with the addition of the Heisenberg spin-spin exchange coupling 
of the $s=1/2$ CuO$_4^{6-}$ (Cu$^{2+}$) centers can be written as follows:
\begin{equation}
		\mathcal{H} 
		= \mathcal{H}_{ch} + \mathcal{H}_{exc} 
		+ \mathcal{H}_{tr}^{(1)} 
		+ \mathcal{H}_{tr}^{(2)} 
		- \mu \sum_i S_{zi}
		.
		\label{eq:Ham0}
\end{equation}
Here, the first term 
\begin{equation}
	\mathcal{H}_{ch} =
	\Delta \sum_i S_{zi}^2 
	+ V \sum_{\left\langle ij\right\rangle} S_{zi} S_{zj} 
\end{equation}
describes the on-site and inter-site nearest-neighbour density-density correlations, respectively, 
so that $\Delta=U/2$, $U$ being the correlation parameter, and $V>0$. 
The sums run over the sites of a 2D square lattice, $\left\langle ij \right\rangle$ means the nearest neighbors.
The second term 
\begin{equation}
		\mathcal{H}_{exc}
		=
		Js^2 \sum_{\langle ij \rangle} \boldsymbol{\sigma}_i \boldsymbol{\sigma}_j 
\end{equation}
is the  antiferromagnetic ($J>0$) Heisenberg exchange coupling for the CuO$_4^{6-}$ centers, 
where $\boldsymbol{\sigma}=P_0 \mathbf{s}/s$ operators take into account the on-site spin density $P_0 = 1-S_z^2$, 
and $\mathbf{s}$ is the spin $s=1/2$ operator \cite{Panov2016}. 
The third term
\begin{multline}
	\mathcal{H}_{tr}^{(1)}
	\;=\;
	- t_p \sum_{\left\langle ij\right\rangle \sigma} 
		\big(  P_{i}^{\sigma{+}} P_{j}^{\sigma}  +   P_{j}^{\sigma{+}} P_{i}^{\sigma} \big)
	\;- \\
	-\; t_n \sum_{\left\langle ij\right\rangle \sigma} 
		\big(  N_{i}^{\sigma{+}} N_{j}^{\sigma}  +  N_{j}^{\sigma{+}} N_{i}^{\sigma}  \big)
	\;- \\
	{} - \frac{1}{2} \, t_{pn} 
	\sum_{\left\langle ij\right\rangle \sigma} 
		\big(  P_{i}^{\sigma{+}} N_{j}^{\sigma}  +  P_{j}^{\sigma{+}} N_{i}^{\sigma} 
		+ N_{i}^{\sigma{+}} P_{j}^{\sigma}  +  N_{j}^{\sigma{+}} P_{i}^{\sigma}  \big)
\end{multline}
with the transfer integrals $t_p$, $t_n$, $t_{pn}$ 
describes the three types of the correlated "one-particle" transport
\cite{Moskvin2011,Moskvin2013}.
$\sigma$ is the spin index, $\sigma=\,\uparrow,\,\downarrow$, 
and the orbital part of the $P$ and $N$ operators expressed in terms of the pseudospin $S{=}1$ operators: 
$P^{+} \propto \left(S_{+} + T_{+}\right)$, $N^{+} \propto \left(S_{+} - T_{+}\right)$, $T_{+} = S_z S_{+} + S_{+} S_z$.
The next term
\begin{equation}
	\mathcal{H}_{tr}^{(2)}
	=	- t_b \sum_{\left\langle ij\right\rangle} \big(  S_{{+}i}^2 S_{{-}j}^2 + S_{{+}j}^2 S_{{-}i}^2  \big)
\end{equation}
with the transfer integral $t_b$ describes the two-particle ("composite boson") transport
\cite{Moskvin2011,Moskvin2013}.
The last term with chemical potential $\mu$ is needed to account for the charge density constraint, 
$nN = \left\langle \sum_i S_{zi} \right\rangle = const$.

In this paper we present the mean-field results for the spin-pseudospin model (\ref{eq:Ham0}). 
We use the simplest Hartree approximation  
and get the explicit form of the equations for the critical temperatures of the most significant order parameters of the model.


\section{Mean-field approximation}

Using the combinations $B_x = S_{-}^2 + S_{+}^2$, $B_y = i\left(S_{-}^2 - S_{+}^2\right)$, 
we write $\mathcal{H}_{tr}^{(2)}$ in the form
\begin{equation}
	\mathcal{H}_{tr}^{(2)} =
	- \frac{t_b}{2} \sum_{\langle ij \rangle} \mathbf{B}_{i} \mathbf{B}_{j}
	.
\end{equation}
To simplify further calculations, we will use the Hartree approximation. 
This allows us to rewrite the expression for $\mathcal{H}_{tr}^{(1)}$ as
\begin{multline}
	\mathcal{H}_{tr}^{(1)} =
	- \frac{t_p}{2} \sum_{\langle ij \rangle \sigma} \mathbf{P}_{i}^{\sigma} \mathbf{P}_{j}^{\sigma} 
	- \frac{t_n}{2} \sum_{\langle ij \rangle \sigma} \mathbf{N}_{i}^{\sigma} \mathbf{N}_{j}^{\sigma} 
	-{}\\
	{}- \frac{t_{pn}}{4} \sum_{\langle ij \rangle \sigma} 
		\left( \mathbf{P}_{i}^{\sigma} \mathbf{N}_{j}^{\sigma} + \mathbf{N}_{i}^{\sigma} \mathbf{P}_{j}^{\sigma} \right)
	,
\end{multline}
where $P_x^{\sigma} = P^{\sigma} + P^{\sigma{+}}$, $P_y^{\sigma} = i \left( P^{\sigma} - P^{\sigma{+}} \right)$, 
$N_x^{\sigma} = N^{\sigma} + N^{\sigma{+}}$, $N_y^{\sigma} = i \left( N^{\sigma} - N^{\sigma{+}} \right)$.

We use the mean-field approximation and the Bogolubov inequality for the grand potential $\Omega(\mathcal{H})$: 
$\Omega(\mathcal{H}) = \Omega(\mathcal{H}_0) + \left\langle \mathcal{H} - \mathcal{H}_0 \right\rangle$,  
to estimate the free energy of a system per one site, $f = \Omega/N + \mu n$.
Within a two-sublattice ($A$ and $B$) approximation, we introduce the Hamiltonian $\mathcal{H}_0$
\begin{equation}
	\mathcal{H}_0 = \sum_{c=1}^{N/2} \mathcal{H}_c 
	,\qquad
	\mathcal{H}_c = \mathcal{H}_A + \mathcal{H}_B
	,
\end{equation}
\begin{multline}
	\mathcal{H}_\alpha = \Delta S_{z\alpha}^2 
	- \left( h_z \pm h_z^a \right) S_{z\alpha} 
	- \left( \mathbf{g} \pm \mathbf{g}^{a} \right) \boldsymbol{\sigma}_{\alpha} 
	- \left( \mathbf{h}_2 \pm \mathbf{h}_2^{a} \right) \mathbf{B}_{\alpha} 
	-{}\\
	{}- \sum_{\sigma} \left( \mathbf{h}_p^{\sigma} \pm \mathbf{h}_p^{a,\sigma} \right) \mathbf{P}_{\alpha}^{\sigma}
	- \sum_{\sigma} \left( \mathbf{h}_n^{\sigma} \pm \mathbf{h}_n^{a,\sigma} \right) \mathbf{N}_{\alpha}^{\sigma}
	,
\end{multline}
where $\alpha=A,B$, the upper (lower) sign corresponds to $A$ ($B$) sublattice, and 
$h_z$, $h_z^a$, $\mathbf{g}$, $\mathbf{g}^{a}$, $\mathbf{h}_2$, $\mathbf{h}_2^{a}$, 
$\mathbf{h}_p^{\sigma}$, $\mathbf{h}_p^{a,\sigma}$, $\mathbf{h}_n^{\sigma}$, $\mathbf{h}_n^{a,\sigma}$ 
($\sigma=\,\uparrow,\,\downarrow$)
are the molecular fields.
Using the partition function $Z_c = \mathrm{Tr\,} \left[ \exp\left( -\beta \mathcal{H}_c \right) \right]$, $\beta=1/T$, 
we obtain the expressions for the charge density $n$ and the order parameters:
\begin{equation}
	n = \frac{1}{2\beta} \frac{\partial \ln Z_c}{\partial h_z}
	,\qquad
	a = \frac{1}{2\beta} \frac{\partial \ln Z_c}{\partial h_z^a}
	,
	\label{eq:PP1}
\end{equation}
\begin{equation}
	\mathbf{m} = \frac{1}{2\beta} \frac{\partial \ln Z_c}{\partial \mathbf{g}}
	,\qquad
	\mathbf{l} = \frac{1}{2\beta} \frac{\partial \ln Z_c}{\partial \mathbf{g}^{a}}
	,
	\label{eq:PP2}
\end{equation}
\begin{equation}
	\mathbf{B} = \frac{1}{2\beta} \frac{\partial \ln Z_c}{\partial \mathbf{h}_2}
	,\qquad
	\mathbf{B}_a = \frac{1}{2\beta} \frac{\partial \ln Z_c}{\partial \mathbf{h}_2^{a}}
	,
	\label{eq:PP3}
\end{equation}
\begin{equation}
	\mathbf{P}^{\sigma} = \frac{1}{2\beta} \frac{\partial \ln Z_c}{\partial \mathbf{h}_p^{\sigma}}
	,\qquad
	\mathbf{P}_a^{\sigma} = \frac{1}{2\beta} \frac{\partial \ln Z_c}{\partial \mathbf{h}_p^{a,\sigma}}
	,
	\label{eq:PP4}
\end{equation}
\begin{equation}
	\mathbf{N}^{\sigma} = \frac{1}{2\beta} \frac{\partial \ln Z_c}{\partial \mathbf{h}_n^{\sigma}}
	,\qquad
	\mathbf{N}_a^{\sigma} = \frac{1}{2\beta} \frac{\partial \ln Z_c}{\partial \mathbf{h}_n^{a,\sigma}}
	.
	\label{eq:PP5}
\end{equation}
The free energy per one site is given by
\begin{multline}
	f = 
	-\frac{1}{2\beta} \ln Z_c 
	+ 2V \left( n^2 - a^2 \right) 
	+{}\\
	{} 
	+ 2Js^2 \left( \mathbf{m}^2 - \mathbf{l}^2 \right) 
	- t_b \left( \mathbf{B}^2 - \mathbf{B}_a^2 \right) 
	-{}\\[0.5em]
	{}	
	- t_p \sum_{\sigma} \left( {\mathbf{P}^{\sigma}}^2 - {\mathbf{P}_a^{\sigma}}^2 \right) 
	- t_n \sum_{\sigma} \left( {\mathbf{N}^{\sigma}}^2 - {\mathbf{N}_a^{\sigma}}^2 \right) 
	-{}\\
	{}	
	- t_{pn} \sum_{\sigma} 
		\left( \mathbf{P}^{\sigma} \mathbf{N}^{\sigma} - \mathbf{P}_a^{\sigma} \mathbf{N}_a^{\sigma} \right) 
	+{}\\
	{}+ 
	h_z n + h_z^a a 
	+ \mathbf{g} \mathbf{m}	+ \mathbf{g}^a \mathbf{l} 
	+ \mathbf{h}_2 \mathbf{B} + \mathbf{h}_2^a \mathbf{B}_a 
	+{}\\[0.5em]
	{} 
	+ \sum_{\sigma} \left( 
	\mathbf{h}_p^{\sigma} \mathbf{P}^{\sigma}
	+ \mathbf{h}_p^{a,\sigma} \mathbf{P}_a^{\sigma}
	+ \mathbf{h}_n^{\sigma} \mathbf{N}^{\sigma}
	+ \mathbf{h}_n^{a,\sigma} \mathbf{N}_a^{\sigma}
	\right)
	.
\end{multline}

By minimizing the free energy, we get a system of equations to determine the values of the order parameters:
\begin{equation}
	4 V a = h_z^a
	,\quad
	- 4 J s^2 \mathbf{m} = \mathbf{g} 
	,\quad
	4 J s^2 \mathbf{l} = \mathbf{g}^a
	,
	\label{eq:orderPsys1}
\end{equation}

\begin{equation}
	2 t_b \mathbf{B} = \mathbf{h}_2
	,\quad
	- 2 t_b \mathbf{B}_a = \mathbf{h}_2^a
	,
	\label{eq:orderPsys2}
\end{equation}
\begin{equation}
	2 t_p \mathbf{P}^{\sigma} + t_{pn} \mathbf{N}^{\sigma} = \mathbf{h}_p^{\sigma}
	,\quad
	t_{pn} \mathbf{P}^{\sigma} + 2 t_n \mathbf{N}^{\sigma} = \mathbf{h}_n^{\sigma}
	,
	\label{eq:orderPsys3}
\end{equation}
\begin{equation}
	- 2 t_p \mathbf{P}_a^{\sigma} - t_{pn} \mathbf{N}_a^{\sigma} = \mathbf{h}_p^{a,\sigma}
	,\quad
	- t_{pn} \mathbf{P}_a^{\sigma} - 2 t_n \mathbf{N}_a^{\sigma} = \mathbf{h}_n^{a,\sigma}
	.
	\label{eq:orderPsys4}
\end{equation}

From the stability conditions for a minimum of $f$ that corresponds to the high-temperature disordered phase, we obtain the equations for the critical temperatures. 
\begin{equation}
	4 V  = \left.\frac{\partial h_z^a}{\partial a}\right|_0
	,
	\label{eq:Tc0}
\end{equation}

\begin{equation}
	- 4 J s^2 = \left.\frac{\partial g_{\alpha}}{\partial m_{\alpha}}\right|_0
	,\quad
	4 J s^2 = \left.\frac{\partial g_{\alpha}^a}{\partial l_{\alpha}}\right|_0
	,\quad
	\alpha = x,y,z;
	\label{eq:Tc1}
\end{equation}

\begin{equation}
	2 t_b = \left.\frac{\partial h_{2,\alpha}}{\partial B_{\alpha}}\right|_0
	,\quad
	- 2 t_b = \left.\frac{\partial h_{2,\alpha}^a}{\partial B_{a,\alpha}}\right|_0
	,\quad
	\alpha = x,y;
	\label{eq:Tc2}
\end{equation}
\begin{multline}
	\left. \frac{ \partial h_{p,\alpha}^{\sigma} }{ \partial P_{\alpha}^{\sigma} } \right|_0 
	+ \left. \frac{ \partial h_{n,\alpha}^{\sigma} }{ \partial N_{\alpha}^{\sigma} } \right|_0 
	- 2 \left( t_p + t_n \right)
	\pm {}\\
	{}
	\pm \sqrt{ 
	\left(
	\left. \frac{ \partial h_{p,\alpha}^{\sigma} }{ \partial P_{\alpha}^{\sigma} } \right|_0 
	- \left. \frac{ \partial h_{n,\alpha}^{\sigma} }{ \partial N_{\alpha}^{\sigma} } \right|_0 
	- 2\left( t_p - t_n \right)
	\right)^2 
	+ 4 t_{pn}^2
	}
	= 0
	,\\
	\alpha = x,y;
	\label{eq:Tc3}
\end{multline}
\begin{multline}
	\left. \frac{ \partial h_{p,\alpha}^{a,\sigma} }{ \partial P_{a,\alpha}^{\sigma} } \right|_0 
	+ \left. \frac{ \partial h_{n,\alpha}^{a,\sigma} }{ \partial N_{a,\alpha}^{\sigma} } \right|_0 
	+ 2 \left( t_p + t_n \right)
	\pm {}\\
	{}
	\pm \sqrt{ 
	\left(
	\left. \frac{ \partial h_{p,\alpha}^{a,\sigma} }{ \partial P_{a,\alpha}^{\sigma} } \right|_0 
	- \left. \frac{ \partial h_{n,\alpha}^{a,\sigma} }{ \partial N_{a,\alpha}^{\sigma} } \right|_0 
	+ 2\left( t_p - t_n \right)
	\right)^2 
	+ 4 t_{pn}^2
	}
	= 0
	,\\
	\alpha = x,y.
	\label{eq:Tc4}
\end{multline}
Here, the index 0 denotes the minimum of the high-temperature disordered phase, 
where all molecular fields are zero, except for $h_z$. 
In equations (\ref{eq:Tc3},\ref{eq:Tc4}) we used that
\begin{equation}
	\left. \frac{\partial h_{p,\alpha}^{\sigma}}{\partial N_{\alpha}^{\sigma}} \right|_0 
	= \left. \frac{\partial h_{n,\alpha}^{\sigma}}{\partial P_{\alpha}^{\sigma}} \right|_0 
	= 0
	,\quad
	\alpha = x,y;
	\label{eq:PNzero}
\end{equation}
\begin{equation}
	\left. \frac{\partial h_{p,\alpha}^{a,\sigma}}{\partial N_{a,\alpha}^{\sigma}} \right|_0 
	= \left. \frac{\partial h_{n,\alpha}^{a,\sigma}}{\partial P_{a,\alpha}^{\sigma}} \right|_0 
	= 0
	,\quad
	\alpha = x,y.
	\label{eq:PNAzero}
\end{equation}
The proof of this statement is given in the Appendix. 
In the case $t_{pn}=0$, the equations (\ref{eq:Tc3},\ref{eq:Tc4}) have the following form
\begin{equation}
	2 t_p = \left. \frac{ \partial h_{p,\alpha}^{\sigma} }{ \partial P_{\alpha}^{\sigma} } \right|_0
	,\quad
	2 t_n = \left. \frac{ \partial h_{n,\alpha}^{\sigma} }{ \partial N_{\alpha}^{\sigma} } \right|_0
	,\quad
	\alpha = x,y
	;
	\label{eq:Tc5}
\end{equation}
\begin{equation}
	- 2 t_p = \left. \frac{ \partial h_{p,\alpha}^{a,\sigma} }{ \partial P_{a,\alpha}^{\sigma} } \right|_0
	,\quad
	- 2 t_n = \left. \frac{ \partial h_{n,\alpha}^{a,\sigma} }{ \partial N_{a,\alpha}^{\sigma} } \right|_0
	,\quad
	\alpha = x,y
	.
\end{equation}


\section{Critical temperatures}

In the high-temperature disordered phase, all molecular fields are zero, except for $h_z$. 
The partition function is given by
\begin{equation}
	Z_{c} = 4 \left(1 + e^{-\delta} \cosh \eta_z\right)^2
	,
\end{equation}
where $\delta=\beta\Delta$, $\eta_z=\beta h_z$. We can find $n$ using (\ref{eq:PP1}), 
and get an explicit expression for the molecular field $h_z$:
\begin{equation}
	\eta_z = \beta h_z 
	= \frac{1}{2} \ln \frac{\left(1 + n\right)\left(\phi + n\right)}{\left(1 - n\right)\left(\phi - n\right)}
	,	
	\label{eq:etaz}
\end{equation}
where $\phi = \sqrt{ \left( 1 - n^2 \right) e^{-2\delta} + n^2 }$.

Given the signs of the $V$, $J$ and the transfer integrals  $t_p$, $t_n$, $t_{pn}$ and $t_b$, non-trivial solutions for the critical temperatures exist for the following order parameters:  
$a$, $\mathbf{l}$, $\mathbf{B}$, $\mathbf{P}^{\sigma}$, $\mathbf{N}^{\sigma}$.
Taking into account the isotropy of the exchange interaction and transfer, 
we write the explicit analytical form of the equations for the critical temperatures of 
$a$, $l_z$, $B_x$, $P_x^{\sigma}$ and $N_x^{\sigma}$.

In order to find the critical temperatures of the charge ordering, $T_{CO}$, we take a small variation of $h_z^a$
at the minimum of the high-temperature disordered phase  
and obtain the partition function to be
\begin{equation}
	Z_{c} = 4 \left(1 + e^{-\delta} \cosh \left( \eta_z + \eta_a \right)\right)
	\left(1 + e^{-\delta} \cosh \left( \eta_z - \eta_a \right)\right)
	,
\end{equation}
where $\eta_a = \beta h_z^a$. This allows us to write the equation (\ref{eq:Tc0}) as
\begin{equation}
	4 \nu  = \frac{ \left( e^\delta + \cosh \eta_z \right)^2 }{ e^\delta \cosh \eta_z  +  1 }
	,
\end{equation}
where $\nu=\beta V$. Using (\ref{eq:etaz}), we obtain the equation for $T_{CO}$ in the form:
\begin{equation}
	4 \nu \left( 1 - n^2 \right) = 1 + \phi^{-1}
	.
	\label{eq:TCO}
\end{equation}
This equation reproduce the result of our recent work \cite{Panov2019}, and generalize the well-known result for the critical temperatures of the charge ordering for the hard-core bosons \cite{Micnas1990}.

The variation of $g_z^a$ 
at the minimum of the high-temperature disordered phase  
to find the equation for the critical temperature of antiferromagnetic ordering, $T_{AFM}$, gives the partition function
\begin{equation}
	Z_{c} = 4 \left( \cosh \gamma_a + e^{-\delta} \cosh \eta_z \right)^2
	,
\end{equation}
where $\gamma_a = \beta g_z^a$. The equation (\ref{eq:Tc1}) for $l_z$ takes the form
\begin{equation}
	4 j  = 1 + e^{-\delta} \cosh \eta_z
\end{equation}
where $j = \beta Js^2$. Using (\ref{eq:etaz}), we obtain the equation for $T_{AFM}$ in the form:
\begin{equation}
	4 j \left( 1 - n^2 \right) = 1 + \phi
	.
\end{equation}
This equation was also found in \cite{Panov2019}.

The variation of $h_2$ 
at the minimum of the high-temperature disordered phase 
yields the partition function
\begin{equation}
	Z_{c} = 4 \left( 1 + e^{-\delta} \cosh \sqrt{\eta_z^2 + \eta_2^2} \right)^2
	,
\end{equation}
where $\eta_2=\beta h_2$. The equation (\ref{eq:Tc1}) for $B_x$ takes the form
\begin{equation}
	2 \tau n = \eta_z
	,
\end{equation}
where $\tau = \beta t_b$, and from (\ref{eq:etaz}) 
we obtain the equation for the superconducting temperature $T_B$ 
(or the critical temperature of superfluidity of the composite bosons) 
\begin{equation}
	4 \tau n = \ln \frac{\left(1 + n\right)\left(\phi + n\right)}{\left(1 - n\right)\left(\phi - n\right)}
	.
\end{equation}
This equation generalizes the well-known result for the superconducting temperature of the preformed local pairs (or the critical temperature of superfluidity of the charged hard-core bosons) \cite{Micnas1990}.

\begin{figure*}
   \includegraphics[width=0.9\textwidth]{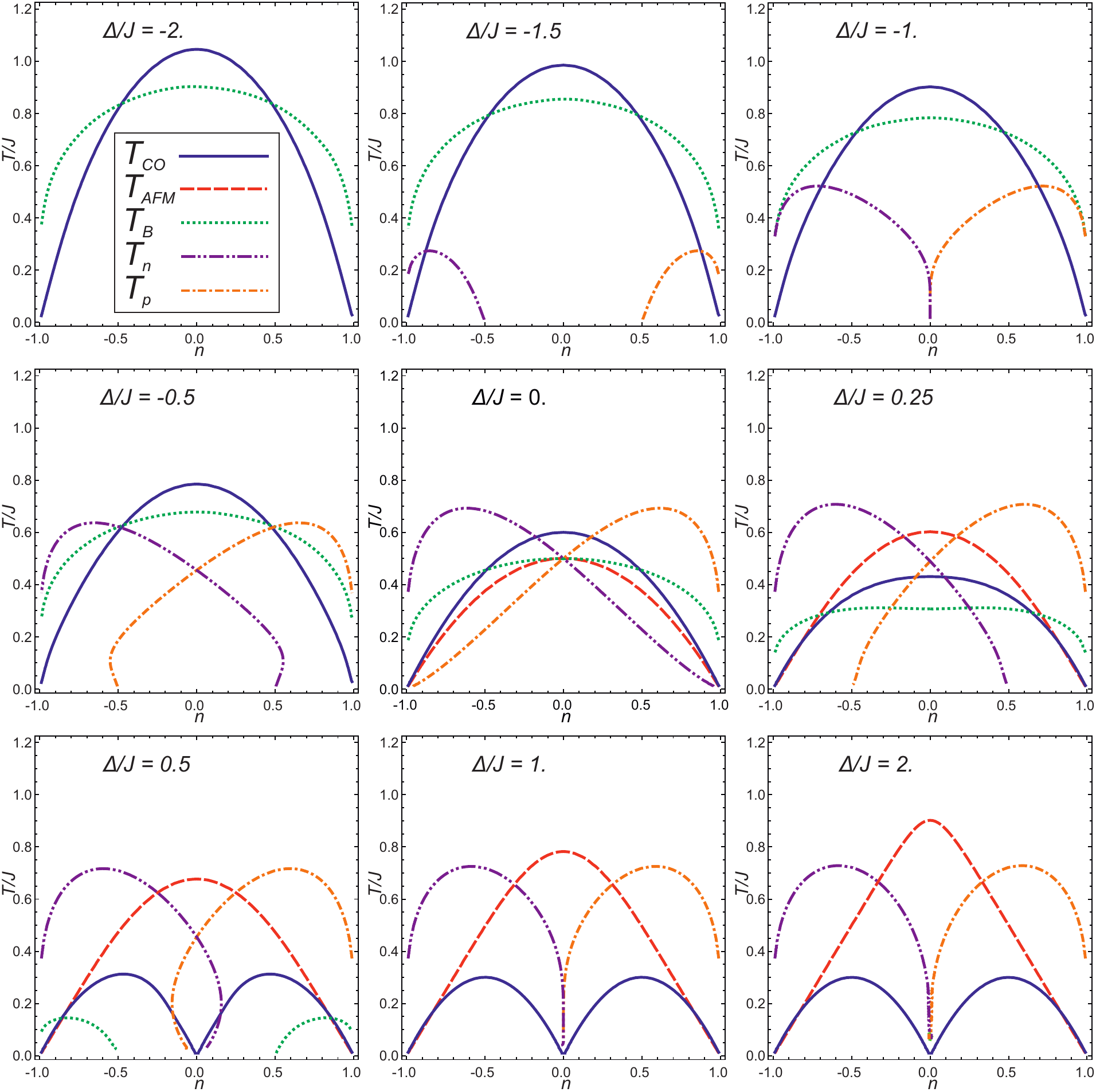} 
   \caption{
	Concentration dependencies of the critical temperatures
	at $V/J=0.3$, $t_B/J=0.5$, $t_p/J=0.5$, $t_n/J=0.5$, $t_{pn}=0$, for different values of $\Delta$.
	}
\label{fig:1}
\end{figure*}

Using the same considerations for the $P_x^{\sigma}$, $\sigma=\uparrow,\downarrow$, we get the partition function
\begin{equation}
	Z_{c} = 4 
	\left(  
		1 + e^{-\delta-\eta_z} + 2 e^{ -\frac{\delta-\eta_z}{2} } 
		\cosh \sqrt{ \left(\frac{\delta-\eta_z}{2}\right)^2 + \eta_p^2 } 
	\right)^2
	,
\end{equation}
where $\eta_p^2 = {\eta_p^{\downarrow}}^2 + {\eta_p^{\uparrow}}^2$, $\eta_p^{\sigma}=\beta h_{p,x}^{\sigma}$.
This yields
\begin{equation}
	\left. \frac{ \partial \eta_{p}^{\sigma} }{ \partial P_{x}^{\sigma} } \right|_0 
	= \frac{\left(1 + e^{-\delta} \cosh \eta_z\right)\left(\delta - \eta_z\right)}{ 1 - e^{-\delta+\eta_z} }
	.
	\label{eq:etaP}
\end{equation} 
Taking into account (\ref{eq:etaz}) and the first equation (\ref{eq:Tc5}), 
we obtain the critical temperature $T_p$ for the case $t_{pn}=0$ in the form
\begin{equation}
	2 \tau_p  = \frac{1+\phi}{\left(1+n\right)\left(1-2n-\phi\right)}  \ln \frac{1 - n}{\phi + n}
	.
\end{equation}
where $\tau_p = \beta t_p$. 
For the order parameters $N_x^{\sigma}$, $\sigma=\uparrow,\downarrow$, we get 
\begin{equation}
	\left. \frac{ \partial \eta_{n}^{\sigma} }{ \partial N_{x}^{\sigma} } \right|_0 
	= \frac{\left(1 + e^{-\delta} \cosh \eta_z\right)\left(\delta + \eta_z\right)}{ 1 - e^{-\delta-\eta_z} }
	,
	\label{eq:etaN}
\end{equation} 
where $\eta_n^{\sigma}=\beta h_{n,x}^{\sigma}$, 
and the second equation (\ref{eq:Tc5}) yields the critical temperature $T_n$ for the case $t_{pn}=0$
\begin{equation}
	2 \tau_n  = \frac{1+\phi}{\left(1-n\right)\left(1+2n-\phi\right)}  \ln \frac{1 + n}{\phi - n}
	,
\end{equation}
where $\tau_n = \beta t_n$. 
If $t_{pn}\neq0$, the critical temperatures are defined by combination of the equations 
(\ref{eq:Tc3}), (\ref{eq:etaz}), (\ref{eq:etaP}) and (\ref{eq:etaN}).

In Fig.\,\ref{fig:1} we reproduce the concentration dependencies of the critical temperatures for different values of the local correlation parameter. 
When $\Delta\to -\infty$, the system associated with (\ref{eq:Ham0}) 
is similar to the charged hard-core bosons \cite{Micnas1990} since the CuO$_4^{6-}$ energy level is high enough. 
Therefore, only two ordered phases are possible: charge-ordered and superconducting.
As $\Delta$ increases, new orderings appear: one is antiferromagnetic, and the other two are associated with one-particle transport of $p$ or $n$ type. 
In the $\Delta\to +\infty$ limit, the superconducting ordering vanishes,  
since the energies of the CuO$_4^{5-,7-}$ states determining the two-particle transport become too high.
In this case, for $n=0$, only one antiferromagnetic ordering is possible, 
while at $n\neq0$ it competes with the charge and $p$ or $n$ orderings, which are ``induced'' by the charge density constraint.
The most difficult situation is realized in a rather narrow range of the $\Delta$ value near zero 
when all ordered phases have comparable energies and compete with each other.

\begin{figure}
	\includegraphics[width=0.45\textwidth]{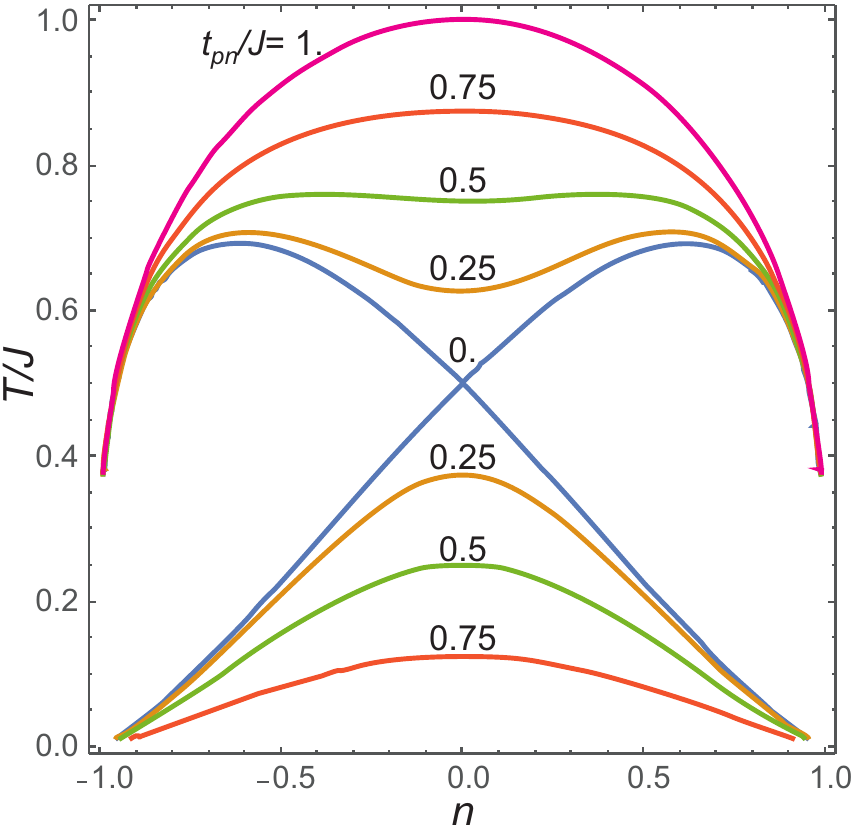}
	\caption{Concentration dependencies of the critical temperatures given by (\ref{eq:Tc3})
	at $\Delta=0$, $t_p/J=0.5$, $t_n/J=0.5$, for different values of $t_{pn}$.
	The lower curves show the solutions of equation (\ref{eq:Tc3}), which has a plus sign, 
	the upper curves show the solutions of the equation with a minus sign.}
	\label{fig:2}
\end{figure}

An additional complication caused by nonzero values of $t_{pn}$ is shown in Fig.\,\ref{fig:2}. 
In this case, there is two modes of mixed $p-n$ one-particle transport, 
that correspond to the solutions of equation (\ref{eq:Tc3}), which has a plus or minus sign. 
When $t_{pn}$ increases, the solutions for the equation with a plus sign decrease 
and disappear at some critical value of $t_{pn}$. 
We also note that equations (\ref{eq:Tc4}) will possess the solutions with similar properties 
starting with some non-zero value of $t_{pn}$.

\medskip
This work was supported by Program 211 of the Government of the Russian Federation (Agreement 02.A03.21.0006),
the Ministry of Education and Science of the Russian Federation (projects nos. 2277 and 5719).


\section*{Appendix}

If the parameter $t_{pn}\neq 0$, then it is necessary to analyze the case of simultaneous variation of $\eta_p^{\sigma}$ and $\eta_n^{\sigma}$.
The molecular fields $\eta_z$, $\eta_p^{\sigma}$, $\eta_n^{\sigma}$ and the order parameters $n$, $P^{\sigma}$ and $N^{\sigma}$ in this case are related by the equations
\begin{equation}
	n = \frac{\partial \ln Z}{\partial \eta_z}
	,\quad
	P^{\sigma} = \frac{\partial \ln Z}{\partial \eta_p^{\sigma}}
	,\quad
	N^{\sigma} = \frac{\partial \ln Z}{\partial \eta_n^{\sigma}}
	,
	\label{eq:nPN}
\end{equation}
where the partition function
\begin{equation}
	Z = \sum_{l=1}^{4} e^{-\lambda_l}
\end{equation}
expressed through the roots $\lambda_l$, $l=1,\ldots4$, of the secular equation
\begin{multline}
	\Phi
	\equiv  \lambda^4 -2\delta\lambda^3 
	+ \lambda^2 \left( \delta^2 - \eta_z^2 - \eta_p^2 - \eta_n^2 \right)
	+{}\\
	+ \lambda \left( \delta \left( \eta_p^2 + \eta_n^2 \right) 
			+ \eta_z \left( \eta_p^2 - \eta_n^2 \right) \right)
	+ \left( \eta_p^{\uparrow} \eta_n^{\uparrow} - \eta_p^{\downarrow} \eta_n^{\downarrow} \right)^2
	= 0	
	.
	\label{eq:sec}
\end{multline}
where $\eta_\alpha^2 = {\eta_\alpha^{\uparrow}}^2 + {\eta_\alpha^{\downarrow}}^2$, $\alpha=p,n$. 
Differentiating the equations (\ref{eq:nPN}) by $P^{\sigma}$, we get the system
\begin{equation}
	\sum_{j} \frac{\partial^2 \ln Z}{\partial \eta_i \partial \eta_j} \frac{\partial \eta_j}{\partial P^{\sigma}} = \delta_{x_i,P^{\sigma}}
	,\;
	\eta_i, \eta_j = \eta_z,\eta_p^{\uparrow},\eta_p^{\downarrow},\eta_n^{\uparrow},\eta_n^{\downarrow}
	,
	\label{eq:sys}
\end{equation}
where $x_i = n,P^{\uparrow},P^{\downarrow},N^{\uparrow},N^{\downarrow}$. 
The matrix elements in (\ref{eq:sys}) are
\begin{equation}
	\frac{\partial^2 \ln Z}{\partial \eta_i \partial \eta_j}
	= - \frac{1}{Z^2} \frac{\partial Z}{\partial \eta_i} \frac{\partial Z}{\partial \eta_j}
	+ \frac{1}{Z} \frac{\partial^2 Z}{\partial \eta_i \partial \eta_j}
	,
\end{equation}
and we will use that
\begin{equation}
	\frac{\partial Z}{\partial \eta_i}
	= -\sum_{l=1}^{4} e^{-\lambda_l} \frac{\partial \lambda_l}{\partial \eta_i}
	,
\end{equation}
\begin{equation}
	\frac{\partial^2 Z}{\partial \eta_i \partial \eta_j}
	= \sum_{l=1}^{4} e^{-\lambda_l} 
	\left( \frac{\partial \lambda_l}{\partial \eta_i} \frac{\partial \lambda_l}{\partial \eta_j}
	- \frac{\partial^2 \lambda_l}{\partial \eta_i \partial \eta_j} \right)
	.
\end{equation}
From (\ref{eq:sec}) we obtain 
\begin{equation}
	\frac{d \Phi}{d\eta_i} = 0 
	= \frac{\partial \Phi}{\partial \eta_i}
	+ \frac{\partial \Phi}{\partial \lambda} \frac{\partial \lambda}{\partial \eta_i}
	,
\end{equation}
and see that 
\begin{equation}
	\left.\frac{\partial \Phi}{\partial \eta_i}\right|_0 = 0
	\; \Rightarrow \;
	\left.\frac{\partial \lambda_l}{\partial \eta_i}\right|_0 = 0
	\;\Rightarrow \;
	\left.\frac{\partial Z}{\partial \eta_i}\right|_0 = 0
	, \; \eta_i=\eta_p^{\sigma},\eta_n^{\sigma}.
\end{equation}
Here point 0 is the minimum of the high-temperature disordered phase, 
where all molecular fields are zero, except for $\eta_z$. 
Next, we find the second derivative using the identity
\begin{multline}
	\frac{d^2 \Phi}{d\eta_i d\eta_j} = 0 
	= 
	\frac{\partial^2 \Phi}{\partial \eta_i \partial \eta_j}
	+ \frac{\partial^2 \Phi}{\partial \lambda \partial \eta_j} \frac{\partial \lambda}{\partial \eta_i}
	{}+\\
	{}
	+ \frac{\partial^2 \Phi}{\partial \lambda \partial \eta_i} \frac{\partial \lambda}{\partial \eta_j}
	+ \frac{\partial^2 \Phi}{\partial \lambda^2} \frac{\partial \lambda}{\partial \eta_i} \frac{\partial \lambda}{\partial \eta_j}
	+ \frac{\partial \Phi}{\partial \lambda}  \frac{\partial^2 \lambda}{\partial \eta_i \partial \eta_j}
	,
\end{multline}
and this yields
\begin{equation}
	\left. \frac{\partial^2 \lambda}{\partial \eta_i \partial \eta_j} \right|_0 = 0
	,\quad \eta_i \neq \eta_j ;\; \eta_i=\eta_p^{\uparrow},\eta_p^{\downarrow},\eta_n^{\uparrow},\eta_n^{\downarrow}
	.
\end{equation}
Finally, we get
\begin{equation}
	\left. \frac{\partial^2 Z}{\partial \eta_i \partial \eta_j} \right|_0 = 0
	,\quad \eta_i \neq \eta_j ;\; \eta_i=\eta_p^{\uparrow},\eta_p^{\downarrow},\eta_n^{\uparrow},\eta_n^{\downarrow}
	.
\end{equation}
It means that the matrix in the system (\ref{eq:sys}) at point 0 is diagonal, 
and the equations (\ref{eq:PNzero}) are satisfied. 
Similar considerations are valid for equations (\ref{eq:PNAzero}).



\begin{thebibliography}{99}
\bibitem{Fradkin2012} 
E. Fradkin, S. A. Kivelson, 
``Ineluctable complexity,'' 
Nature Physics \textbf{8}, 864-866 (2012).
\bibitem{Moskvin2011} 
A.S. Moskvin, 
``True charge-transfer gap in parent insulating cuprates,'' 
Phys. Rev. B \textbf{84}, 075116 (2011).
\bibitem{Moskvin2013} 
A.S. Moskvin, 
``Perspectives of disproportionation driven superconductivity in strongly correlated 3d compounds,'' 
J. Phys.: Condens. Matter \textbf{25}, 085601 (2013).
\bibitem{Panov2016}  
Y.D. Panov, A.S. Moskvin, A.A. Chikov, I.L. Avvakumov, 
``Competition of Spin and Charge Orders in a Model Cuprate,''
J. Low Temp. Phys. \textbf{185}, 409 (2016).
\bibitem{Panov2019} 
Y.D. Panov, V.A. Ulitko, K.S. Budrin, A.A. Chikov, A.S. Moskvin, 
``Phase diagrams of a 2D Ising spin-pseudospin model,''
J. Magn. Magn. Mater. \textbf{477}, 162 (2019).
\bibitem{Micnas1990}  
R. Micnas, J. Ranninger, S. Robaszkiewicz, 
``Superconductivity in narrow-band systems with local nonretarded attractive interactions,''
Rev. Mod. Phys. \textbf{62}, 113 (1990).
\end{thebibliography}
\end{document}